\documentclass[conference]{IEEEtran}
\IEEEoverridecommandlockouts
\usepackage{cite}
\usepackage{amsmath,amssymb,amsfonts}
\usepackage{graphicx}
\usepackage{textcomp}
\usepackage{xcolor}
\def\BibTeX{{\rm B\kern-.05em{\sc i\kern-.025em b}\kern-.08em
		T\kern-.1667em\lower.7ex\hbox{E}\kern-.125emX}}

\usepackage{graphicx,cite,calc,color,subfigmat,amssymb,amsmath,dsfont,epstopdf,tikz}
\usepackage{array}
\usepackage{multirow}
\usepackage{cleveref}

\usepackage{subfigure}
\usepackage{graphicx}
\usepackage{amsmath,amssymb,amsfonts,bm}
\usepackage{breqn}

\usepackage{algorithm}
\usepackage[noend]{algpseudocode}

\makeatletter
\def\BState{\State\hskip-\ALG@thistlm}
\makeatother

\usepackage{graphicx}
\usepackage{caption}

\usepackage{tabularx}
\usepackage{dblfloatfix}

\usepackage{fancyhdr}
\usepackage{kantlipsum}

\usepackage{etoolbox}

\begin{document}
	
	\title{Fluid segmentation in Neutrosophic domain\\
	}
	
	\author{\IEEEauthorblockN{Elyas Rashno}
		\IEEEauthorblockA{\textit{Department of Computer Engineering} \\
			\textit{Iran University Of Science and Technology}\\
			Tehran, Iran \\
			elyas.rashno@gmail.com}
		\and
		\IEEEauthorblockN{Abdolreza Rashno}
		\IEEEauthorblockA{\textit{Department of Computer Engineering } \\
			\textit{Engineering Faculty}\\
			\textit{Lorestan University}\\
			Khorramabad, Iran \\
			rashno.a@lu.ac.ir}
		\and
		\IEEEauthorblockN{Sadegh Fadaei}
		\IEEEauthorblockA{\textit{Department of Electrical Engineering } \\
			\textit{Faculty of Engineering}\\
			\textit{Yasouj University}\\
			Yasouj, Iran, 7591874934 \\
			s.fadaei@yu.ac.ir}

	}
	
	\maketitle

\begin{abstract}

Optical coherence tomography (OCT) as retina imaging technology is currently used by ophthalmologist as a non-invasive and non-contact method for diagnosis of age-related degeneration (AMD) and diabetic macular edema (DME) diseases. Fluid regions in OCT images reveal the main signs of AMD and DME. In this paper, an efficient and fast clustering in neutrosophic (NS) domain referred as neutrosophic C-means is adapted for fluid segmentation.  For this task, a NCM cost function in NS domain is adapted for fluid segmentation and then optimized by gradient descend methods which leads to binary segmentation of OCT Bscans to fluid and tissue regions. The proposed method is evaluated in OCT datasets of subjects with DME abnormalities. Results showed that the proposed method outperforms existing fluid segmentation methods by 6\% in dice coefficient and sensitivity criteria.     
\end{abstract}


\section{Introduction}
Ophthalmologists use optical coherence tomography (OCT) as a non-contact method to diagnosis and follow up of retina disease. Fluid regions are detectable in OCT images and reveal abnormalities of retina disease including diabetic macula edema (DME) and age-related macular degeneration (AMD) \cite{r1}. Image processing and data mining applications have been widely used in many application \cite{rashno2017effectiveNeuCom,rashno2013highlySpeVer,rashno2015textSpeVerConf,rashno2015MarsConf,rashno2014imageImgRes,rashno2014regularizationImgRes}. Fluid regions can be segmented automatically by image processing and data mining tools. Many fluid detection methods have been proposed in the literature such as kernel regression based segmentation of OCT images with DME \cite{r2}, fluid-filled region delineation of boundaries in OCT \cite{r3}, semi-automatic segmentation method for retinal OCT images tested in patients with diabetic macular edema \cite{r4}, label propagation and higher-order constraint-based segmentation of fluid regions in OCT images of DME subjects \cite{r5}, computerized assessment of intraretinal and subretinal fluid regions in OCT images of the retina \cite{r6}, automated segmentation of pathological cavities in OCT scans \cite{r7}, three-dimensional analysis of retinal layer and fluid-filled regions in OCT of the macula \cite{r8}, probability constrained graph-search-graph-cut method for three-dimensional segmentation of fluid-associated abnormalities \cite{r9}, fluid segmentation with shortest-path methods and neutrosophic (NS) theory \cite{r10,r11,r12,r13,r14,r15}, fluid segmentation with deep convolutional networks (CNNs) \cite{r16,r17,r18,r19}.

Smarandache proposed NS theory in 1995, a branch of philosophy which studies the origin, nature and scope of neutralities, as well as their interactions with different ideational spectra \cite{r20}. This theory has been applied on many applications including image segmentation \cite{r21,r22,r23,r24}, image thresholding \cite{r25}, image edge detection \cite{r26}, content-based image retrieval \cite{r27,r28,r29,r30,CBIRSpeedUp2019}, retinal image analysis \cite{r10,r11,r12,r13,r14}, speech processing \cite{r31}, data clustering \cite{r32,r33} and uncertainty handling \cite{r34,r35}. 
The main contribution of this work is to optimize a cost function in NS domain to binarize retina OCT scans to fluid and tissue regions. The cost function is derived from fuzzy c-means (FCM) and is optimized by gradient descend methods in subsequence iterations. The rest of this paper is organized as follows: Section II reviews FCM and NS theory. Section III presents the proposed method. Experimental results and conclusion are discussed in sections IV and V, respectively.

\section{Review of NS and FCM}
Since the cost function in this work is an extension of FCM and is defined in NS domain, here, a review on FCM and NS is discussed shortly.

\subsection{FCM}
The FCM is a clustering method that assigns a membership degree in interval [0-1] to each data point, Therefore, each data point is assigned to all clusters with different membership degrees and the sum of memberships to all clusters should be 1. FCM cost function is defined as follows:
\begin{equation}
J=\sum_{j=1}^{N}\sum_{i=1}^{C}u_{ij}^m||X_j-v_i||^2 \qquad \qquad \qquad \qquad \qquad \qquad
\label{eq1}
\end{equation}

where $ u_{ij} $ represents the membership degree of data point $ X_j $ to the $ i^{th} $ cluster with center $ v_i $. $ ||\cdot|| $ and $ m $  are a norm metric and a constant variable for determining the fuzziness of the resulting partition, respectively. Minimizing FCM cost function leads to the following equations for the computation of membership degrees and cluster centers:
\begin{equation}
u_{ij}=\frac{1}{\displaystyle{\sum_{j=1}^{N}\left(\frac{||X_j-v_i||}{||X_j-v_k}\right)^{\frac{2}{m-1}}}} \qquad \qquad \qquad \qquad \qquad
\label{eq2}
\end{equation}
\begin{equation}
v_i=\frac{\sum_{j=1}^{N}u_{ij}^mX_j}{\sum_{j=1}^{N}u_{ij}^m} \qquad \qquad \qquad \qquad \qquad \qquad \qquad
\label{eq3}
\end{equation}

Firstly, membership degrees are initialized randomly and cluster centers are computed by Eq. (\ref{eq3}). Then, membership degrees and cluster centers are computed repeatedly, until there is no significant changes of these parameters in subsequent iterations.

\subsection{Neutrosophic}
NS theory is a newly brought up branches of philosophy. In this theory, set A, Anti-A (contemplates A in a correlation with its opposite), Neut-A (neutrality of A, neither A nor Anti-A) are defined \cite{r20}.  In each application, NS sets should be defined for all data points as well as relations between sets. For example, in our problem of interest as image segmentation, each pixel in NS domain is modeled as $ P(t,i,f) $ meaning that it is $ t $ percent true, $ i $ percent indeterministic, and $ f $ percent false. Therefore, three sets $ P_{NS}(i,j)=\{T(i,j), I(i,j), F(i,j)\} $ are defined for pixels in NS domain  as follows \cite{r21}:
\begin{equation}
T(i,j)=\frac{\bar{g}(i,j)-\bar{g}_{min}}{\bar{g}_{max}-\bar{g}_{min}} \qquad \qquad \qquad \qquad \qquad \qquad
\label{eq4}
\end{equation}
\begin{equation}
\bar{g}(i,j)=\frac{1}{w \times w} \sum_{m=i-\frac{w}{2}}^{i+\frac{w}{2}}\sum_{n=j-\frac{w}{2}}^{j+\frac{w}{2}} g(m,n) \qquad \qquad \qquad
\label{eq5}
\end{equation}
\begin{equation}
I(i,j)=\frac{\delta(i,j)-\delta_{min}}{\delta_{max}-\delta_{min}} \qquad \qquad \qquad \qquad \qquad \qquad
\label{eq6}
\end{equation}
\begin{equation}
\delta(i,j)=|g(i,j)-\bar{g}(i,j)|  \qquad \qquad \qquad \qquad \qquad \quad
\label{eq7}
\end{equation}
\begin{equation}
F(i,j)=1-T(i,j)   \qquad \qquad \qquad \qquad \qquad \qquad \quad
\label{eq8}
\end{equation}
where, $ g(i,j) $ represents the intensity value of the pixel $ P(i,j) $, $ \bar{g} $ is image matrix $ g $ filtered by average filter with window size $ w $. Difference of two matrixes $ g $ and $ \bar{g} $ is computed and stored in $ \delta $.

\section{Proposed Method}
In this work, NCM cost function is used and adapted for OCT fluid segmentation. Since OCT scans contain noise and there is ambiguity between fluid pixels and many pixels in background and tissue, indeterminacy set $ I $ in NS domain can be very useful and models such ambiguities easily.  Here, each pixel in OCT scans are transferred to three sets $ T $, $ I $ and $ F $ by Eqs. (\ref{eq4}), (\ref{eq6}) and (\ref{eq8}), respectively. Then, NS sets are presented for clustering cost function as follows:
\begin{align}
L(T,I,F,C)&=\sum_{j=1}^{N}\sum_{i=1}^{C}(\bar{w}_1 T_{ij})^m||X_i-C_j||^2 \nonumber \\
&+\sum_{i=1}^{N}(\bar{w}_2 I_i)^m ||X_i-\bar{C}_{i_{max}}||^2 + \sum_{i=1}^{N} \delta^2 (\bar{w}_3 F_i)^m
\label{eq9}
\end{align}
\begin{equation}
\bar{C}_{i_{max}}=\frac{C_{p_i}+C_{q_i}}{2} \qquad \qquad \qquad \qquad \qquad \qquad \qquad
\label{eq10}
\end{equation}
\begin{equation}
C_{p_i}=\text{argmax}(T_{ij}) \quad , \quad j=1,2,\cdots ,C \qquad \qquad \quad
\label{eq11} 
\end{equation}
\begin{equation}
C_{q_i}=\text{argmax}(T_{ij}) \quad , \quad j \neq p_i \cap j=1,2,\cdots ,C \qquad
\label{eq12} 
\end{equation}

Gradient descent method is used for cost function optimization which leads to the following relations for the computation of $ T $, $ I $ and $ F $ sets in NS domain and cluster centers. 

\begin{equation}
T_{ij}=\frac{K}{\bar{w}_1}(X_i-C_j)^{-(\frac{2}{m}-1)} \qquad \qquad \qquad \qquad \quad
\label{eq13}
\end{equation}
\begin{equation}
I_i=\frac{K}{\bar{w}_2}(X_i-\bar{C}_{i_{max}})^{-(\frac{2}{m}-1)} \qquad \qquad \qquad \qquad
\label{eq14}
\end{equation}
\begin{equation}
F_i=\frac{K}{\bar{w}_3} \delta^{-(\frac{2}{m}-1)} \qquad \qquad \qquad \qquad \qquad \qquad \quad
\label{eq15}
\end{equation}
\begin{equation}
C_j=\frac{\sum_{i=1}^{N}(\bar{w}_1 T_{ij})^m X_i}{\sum_{i=1}^{N}(\bar{w}_1 T_{ij})^m} \qquad \qquad \qquad \qquad \qquad
\label{eq16}
\end{equation}
\begin{align}
K&=\Bigg[\frac{1}{\bar{w_1}} \sum_{j=1}^{C}(X_i-C_j)^{-(\frac{2}{m}-1)} \nonumber \\
&+\frac{1}{\bar{w}_2}(X_i-\bar{C}_{i_{max}})^{-(\frac{2}{m}-1)}+\frac{1}{\bar{w}_3} \delta^{-(\frac{2}{m}-1)}\Bigg]^{-1}
\label{eq17}
\end{align}

Parameter $ K $ is a common part in Eqs. (\ref{eq13})-(\ref{eq15}). It is computed once and used three times which leads to speedup in each iteration of optimization.  In the first step, $ 12 $ clusters are considered for clustering. The reason is that there is $ 11 $ layers in retina and each layer can be assigned to one cluster based on its texture and gray level. One extra cluster is considered for fluid regions as fluid cluster. After convergence, clusters are sorted ascendingly based on the gray level of cluster centers. The first cluster is considered as fluid cluster with label 1 and other clusters are considered as tissue clusters with label $ 0 $. Therefore, the proposed clustering scheme leads to a binary segmentation of OCT scans.

\section{Experimental Results}

\subsection{Evaluation Metrics}
To show the effectiveness of the proposed method, it has been evaluated with three metrics; dice coefficient, precision and sensitivity; computed from true positive (TP) fluid pixels detected as fluid, false positive (FP) tissue pixels detected as fluid, true negative (TN) tissue pixels detected as tissue and false negative (FN) fluid pixels detected as tissue.
\begin{equation}
\text{Dice\_Coeff}=\frac{2TP}{2TP+FP+FN} \qquad \qquad \qquad \qquad
\label{eq18}
\end{equation}
\begin{equation}
\text{Sensitivity}=\frac{TP}{TP+FN} \qquad \qquad \qquad \qquad \qquad  \qquad
\label{eq19}
\end{equation}
\begin{equation}
\text{Precision}=\frac{TP}{TP+FP} \qquad \qquad \qquad \qquad \qquad  \qquad
\label{eq20}
\end{equation}

These criteria are used to evaluate fluid segmentation results of automated methods in comparison with manually segmented fluid regions by ophthalmologist experts.

\subsection{Dataset}
The proposed fluid segmentation method is tested on a dataset from OPTIMA Cyst Segmentation Challenge which contains $ 4 $ subjects with $ 49 $ images per subject where the image resolution varies from $ 512 \times 496 $ to $ 512 \times 1024 $. The fluid regions of each OCT image are manually segmented by two ophthalmologist experts as ground truth images. This dataset is publicly available and can be found online\footnote[1]{https://optima.meduniwien.ac.at/research/challenges/}.

\subsection{Resultes}
\begin{figure}[h]
	\centering
	\captionsetup{justification=centering}
	\includegraphics[width= 0.5 \textwidth ]{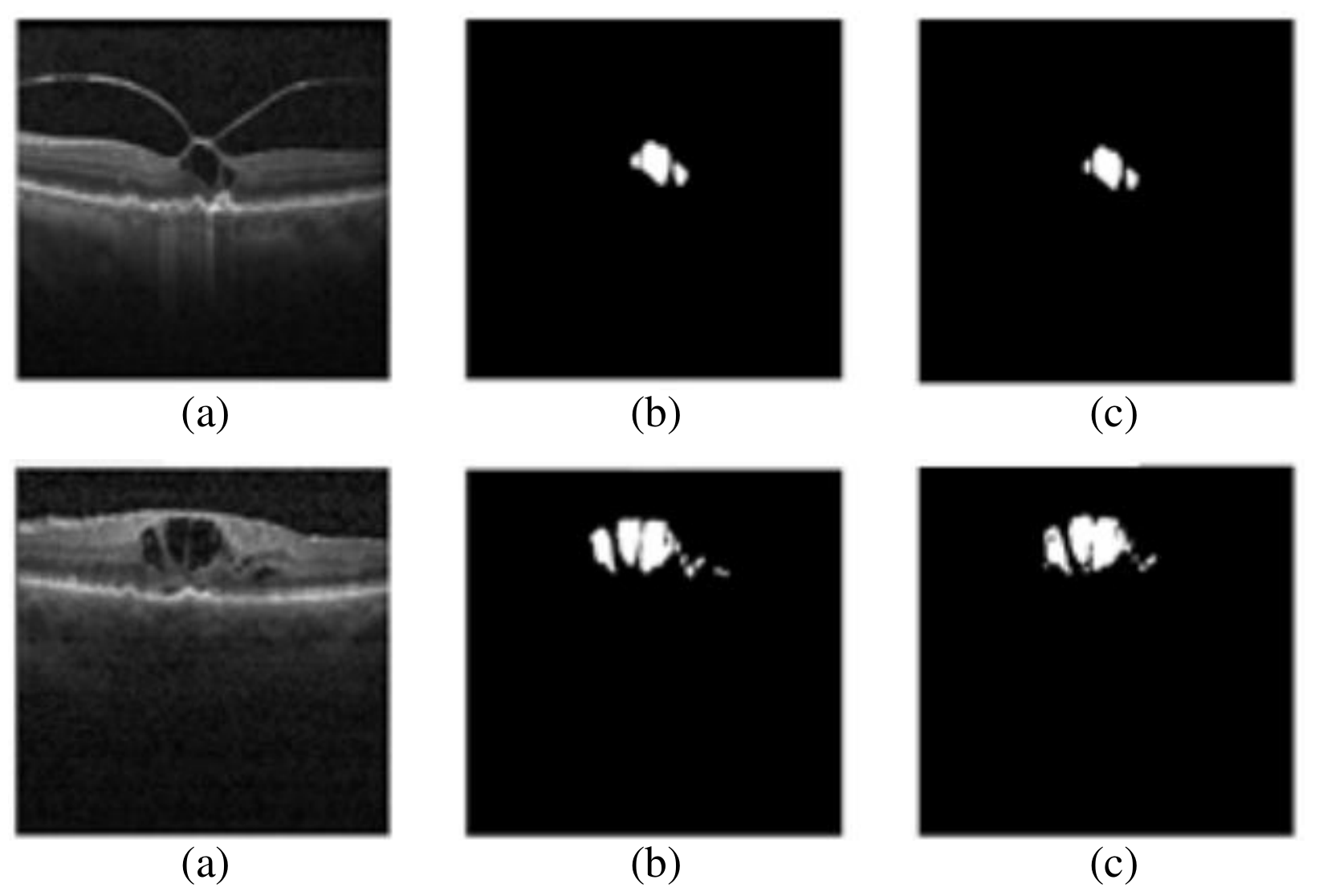}
	\caption{Fluid segmentation results for 2 scans: (a) Input OCT image, (b) manually segmentated by expert and (c) segmented by the proposed method.}
	\label{fig1}
\end{figure}

\begin{figure}[h]
 	\centering
 	\captionsetup{justification=centering}
 	\includegraphics[width= 0.5 \textwidth ]{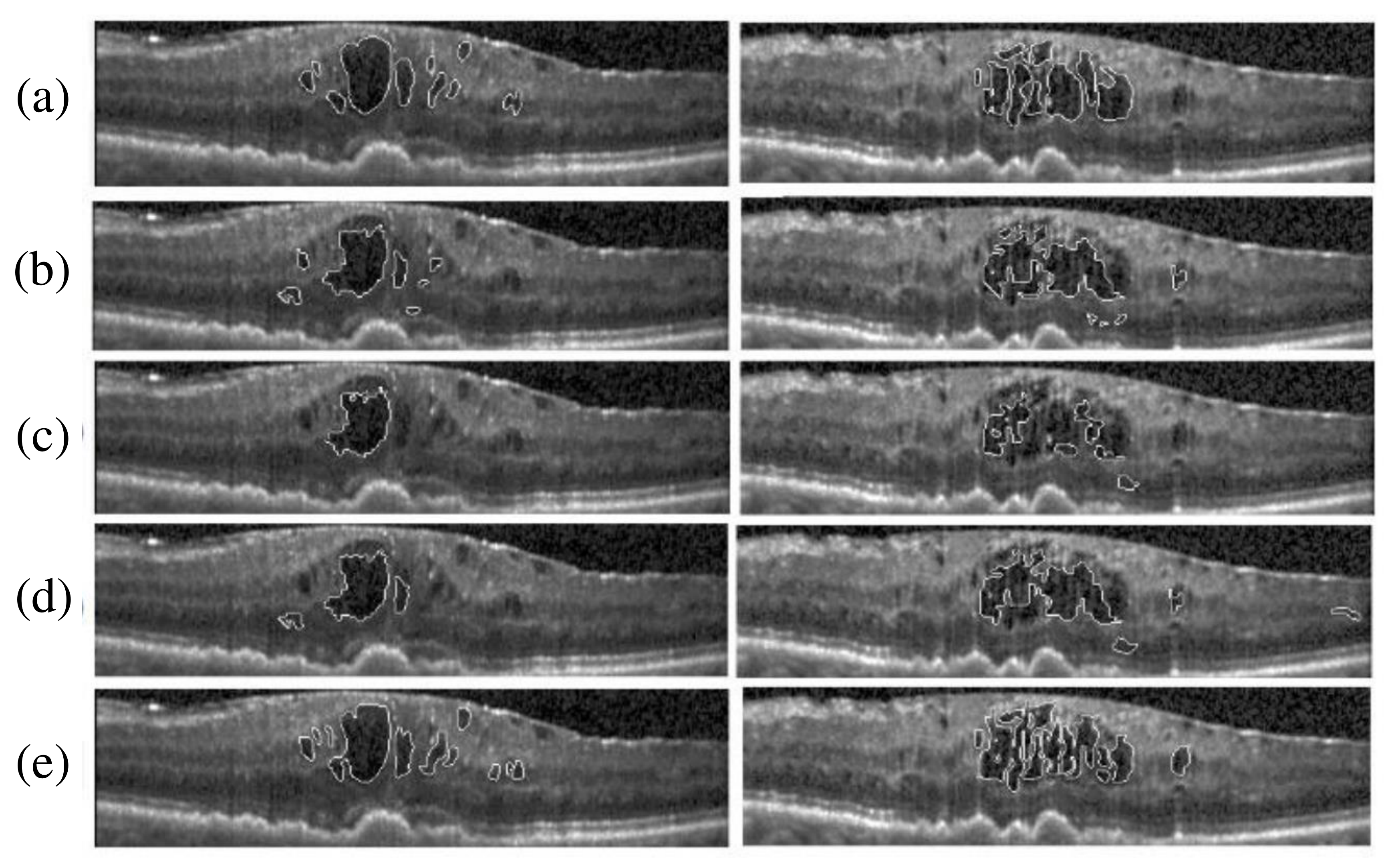}
 	\caption{Segmented fluid by: (a) expert, (b) method in \cite{r11}, (c) method in \cite{r37}, (d) method in \cite{r38} and (e) the proposed method}
 	\label{fig2}
 \end{figure}

Segmented fluid regions by the proposed method are compared with ground truth images. Fig. \ref{fig1} depicts fluid regions segmented by expert and the proposed method for two scans, each scan in one row. The proposed fluid segmentation method is compared with 4 fluid segmentation methods proposed in \cite{r11} and \cite{r37,r38,r39}. The results of these methods are shown in Fig. \ref{fig2}. It is clear visually that the proposed method has lower number of false negaives which leads to higher accuracy.

Table \ref{Table1} reports the dice coefficient, sensitivity and precision of the proposed method and methods in \cite{r11} and \cite{r37,r38,r39}. Reported results in Table \ref{Table1} are the comparison between all methods and fluid regions manually segmented by expert 1. Table \ref{Table2} reports the same results but in comparison with expert 2.  In comparison with the results of expert 1, the proposed method outperforms other methods with respect to dice coefficient and sensitivity of $ 82.23\% $ and $ 90.59\% $, respectively. These measures of the proposed method in comparison with expert 2 are $ 81.62\% $ and $ 90.11\% $ which also outperforms other methods. For precision measure, methods in \cite{r11} with $ 85.29\% $ and \cite{r37} with $ 90.87\% $ have the best performance in comparison with expert 1 and 2, respectively.

\begin{table}[]
	\footnotesize
	\centering
	\caption{Dice coefficient, sensitivity and precision of the proposed method and other methods in comparison with expert 1.}
	\label{Table1}
	\begin{tabular}{|l|l|l|l|l|l|l|}
		\hline
		Expert 1                                          & Sub. & \begin{tabular}[c]{@{}l@{}}Method\\ in {\cite{r37}}\end{tabular} & \begin{tabular}[c]{@{}l@{}}Method\\ in {\cite{r38}}\end{tabular} & \begin{tabular}[c]{@{}l@{}}Method\\ in {\cite{r39}}\end{tabular} & \begin{tabular}[c]{@{}l@{}}Method\\ in {\cite{r11}}\end{tabular} & \begin{tabular}[c]{@{}l@{}}Prop.\\ Method\end{tabular} \\ \hline
		\multicolumn{1}{|c|}{\multirow{5}{*}{Dice Coeff}} & 1    & 73.49                                                        & 80.43                                                        & 71.40                                                        & 82.96                                                        & \textbf{83.17}                                            \\ \cline{2-7} 
		\multicolumn{1}{|c|}{}                            & 2    & 73.90                                                        & 55.10                                                        & 45.49                                                        & \textbf{78.11}                                               & 74.61                                                     \\ \cline{2-7} 
		\multicolumn{1}{|c|}{}                            & 3    & 78.46                                                        & 75.35                                                        & 69.54                                                        & 82.23                                                        & \textbf{84.85}                                            \\ \cline{2-7} 
		\multicolumn{1}{|c|}{}                            & 4    & 78.12                                                        & 71.78                                                        & 71.15                                                        & 80.75                                                        & \textbf{86.32}                                            \\ \cline{2-7} 
		\multicolumn{1}{|c|}{}                            & Ave. & 75.99                                                        & 70.66                                                        & 64.39                                                        & 81.01                                                        & \textbf{82.23}                                            \\ \hline
		\multirow{5}{*}{Sensitivity}                      & 1    & 70.81                                                        & 82.19                                                        & 72.49                                                        & 84.43                                                        & \textbf{86.95}                                            \\ \cline{2-7} 
		& 2    & 96.79                                                        & \textbf{99.04}                                               & 70.45                                                        & 98.94                                                        & 98.42                                                     \\ \cline{2-7} 
		& 3    & 75.72                                                        & 85.13                                                        & 47.38                                                        & 85.18                                                        & \textbf{86.58}                                            \\ \cline{2-7} 
		& 4    & 78.78                                                        & 80.59                                                        & 77.79                                                        & 84.49                                                        & \textbf{90.51}                                            \\ \cline{2-7} 
		& Ave. & 80.52                                                        & 86.73                                                        & 67.02                                                        & 88.26                                                        & \textbf{90.59}                                            \\ \hline
		\multirow{5}{*}{Precision}                        & 1    & \textbf{93.00}                                               & 85.06                                                        & 54.87                                                        & 84.03                                                        & 82.12                                                     \\ \cline{2-7} 
		& 2    & 74.36                                                        & 54.18                                                        & 51.12                                                        & \textbf{78.48}                                               & 69.31                                                     \\ \cline{2-7} 
		& 3    & \textbf{94.89}                                               & 79.88                                                        & 30.93                                                        & 85.45                                                        & 83.43                                                     \\ \cline{2-7} 
		& 4    & \textbf{96.97}                                               & 88.62                                                        & 54.98                                                        & 93.20                                                        & 75.01                                                     \\ \cline{2-7} 
		& Ave. & \textbf{89.80}                                               & 76.93                                                        & 47.97                                                        & 85.29                                                        & 77.46                                                     \\ \hline
	\end{tabular}
\end{table}

\begin{table}[]
	\footnotesize
	\centering
	\caption{Dice coefficient, sensitivity and precision of the proposed method and other methods in comparison with expert 1.}
	\label{Table2}
	\begin{tabular}{|l|l|l|l|l|l|l|}
		\hline
		Expert 2                                          & Sub. & \begin{tabular}[c]{@{}l@{}}Method\\ in {\cite{r37}}\end{tabular} & \begin{tabular}[c]{@{}l@{}}Method\\ in {\cite{r38}}\end{tabular} & \begin{tabular}[c]{@{}l@{}}Method\\ in {\cite{r39}}\end{tabular} & \begin{tabular}[c]{@{}l@{}}Method\\ in {\cite{r11}}\end{tabular} & \begin{tabular}[c]{@{}l@{}}Prop.\\ Method\end{tabular} \\ \hline
		\multicolumn{1}{|c|}{\multirow{5}{*}{Dice Coeff}} & 1    & 72.96                                                        & 79.10                                                        & 68.17                                                        & 82.90                                                        & \textbf{83.03}                                            \\ \cline{2-7} 
		\multicolumn{1}{|c|}{}                            & 2    & 71.68                                                        & 55.11                                                        & 45.81                                                        & \textbf{79.09}                                               & 78.12                                                     \\ \cline{2-7} 
		\multicolumn{1}{|c|}{}                            & 3    & \textbf{82.33}                                               & 79.34                                                        & 65.01                                                        & 80.36                                                        & 82.22                                                     \\ \cline{2-7} 
		\multicolumn{1}{|c|}{}                            & 4    & 77.91                                                        & 71.56                                                        & 72.55                                                        & 80.87                                                        & \textbf{83.12}                                            \\ \cline{2-7} 
		\multicolumn{1}{|c|}{}                            & Ave. & 76.22                                                        & 71.27                                                        & 62.88                                                        & 80.80                                                        & \textbf{81.62}                                            \\ \hline
		\multirow{5}{*}{Sensitivity}                      & 1    & 69.95                                                        & 78.56                                                        & 66.75                                                        & 80.94                                                        & \textbf{83.35}                                            \\ \cline{2-7} 
		& 2    & 92.25                                                        & \textbf{94.54}                                               & 64.71                                                        & 94.45                                                        & 93.86                                                     \\ \cline{2-7} 
		& 3    & 81.49                                                        & 90.95                                                        & 54.84                                                        & 90.75                                                        & \textbf{92.15}                                            \\ \cline{2-7} 
		& 4    & 78.54                                                        & 80.22                                                        & 77.56                                                        & 83.70                                                        & \textbf{91.06}                                            \\ \cline{2-7} 
		& Ave. & 80.55                                                        & 86.06                                                        & 65.96                                                        & 87.46                                                        & \textbf{90.10}                                            \\ \hline
		\multirow{5}{*}{Precision}                        & 1    & \textbf{95.71}                                               & 86.55                                                        & 59.61                                                        & 87.53                                                        & 82.62                                                     \\ \cline{2-7} 
		& 2    & 74.45                                                        & 54.14                                                        & 51.34                                                        & \textbf{78.58}                                               & 67.79                                                     \\ \cline{2-7} 
		& 3    & \textbf{96.10}                                               & 79.96                                                        & 37.99                                                        & 85.48                                                        & 83.14                                                     \\ \cline{2-7} 
		& 4    & \textbf{97.24}                                               & 88.99                                                        & 59.50                                                        & 93.58                                                        & 77.98                                                     \\ \cline{2-7} 
		& Ave. & \textbf{90.87}                                               & 77.41                                                        & 52.11                                                        & 86.29                                                        & 77.88                                                     \\ \hline
	\end{tabular}
\end{table}

\section{Conclusion}

This paper presented a fluid segmentation method based on NCM cost function in NS domain. Minimizing the cost function resulted in binary segmentation of Oct images into tissue and fluid regions. Segmentation results on Optima dataset showed that the proposed method outperforms other segmentation methods in dice coefficient and sensitivity measures while for precision criteria, other methods had the best performance. Future efforts will be directed towards proposing a cost function in NS domain for better modeling of noise and uncertainty in OCT pixels. NS can model uncertainty of OCT pixels in deep convolutional networks and leads to more robust network against noise and imaging device. Therefore, using NS theory for fluid segmentation by CNN can be proposed as another future work.

\end{document}